# Scaling laws in machine learning on non-ideal experimental data enable inverse design of radiative-cooling coatings

Kunihiko Shizume[1,3], Jiawen Li[2], Yaerim Lee[1], Yuki Asano[6], Tuwei Liao[1], Koji Tsuda[2,4,5,*], Junichiro Shiomi[1,4,6,*]

[1] Department of Mechanical Engineering, The University of Tokyo, Tokyo, Japan

[2] Graduate School of Frontier Sciences, The University of Tokyo, Chiba, Japan

[3] Department of Physical Science and Engineering, Nagoya Institute of Technology, Aichi, Japan

[4] RIKEN Center for Advanced Intelligence Project, Tokyo, Japan

[5] Center for Basic Research on Materials, National Institute for Materials Science, Ibaraki, Japan

[6] Institute of Engineering Innovation, The University of Tokyo, Tokyo, Japan

*Correspondence to: Koji Tsuda (tsuda@k.u-tokyo.ac.jp) and Junichiro Shiomi (shiomi@photon.t.u-tokyo.ac.jp)

## Abstract

Materials informatics increasingly targets multicomponent materials, yet experimental datasets inevitably contain non-ideal variations arising from imperfect uniformity and reproducibility in composition, coating formation, and measurement. Whether machine-learning prediction accuracy scales with amount of data under these conditions remains unclear. Here, we develop an automated platform that produced and optically characterized 11,520 radiative-cooling coatings spanning nine material and processing parameters. Forward and inverse neural networks, and a spectrum-conditioned diffusion model were trained on subsets of 10 to 10,000 samples. All models exhibited power-law scaling, with prediction errors decreasing with dataset size. Experimental inverse-design validation showed that larger datasets produced spectra closer to a broadband-emitter target. The diffusion model generated multiple candidate parameter sets, allowing selection among alternative solutions based on ease of fabrication, cost, and scalability. These results demonstrate predictable learning-curve scaling in non-ideal experimental data and provide a quantitative basis for estimating the benefit of additional experimental data, and thus the cost-effectiveness of materials-informatics.

## 1. Introduction

Materials informatics (MI), which integrates machine learning with materials data, has accelerated the discovery of novel materials for applications such as batteries[1], catalysts[2], and thermoelectrics[3]. This progress has been driven by large materials databases such as the Materials Project (~170,000 entries)[4], OQMD (~1,200,000 entries)[5], and AFLOW (~3,500,000 entries)[6], which include both experimentally known and computationally predicted crystal structures and material property data, and has been further amplified by artificial intelligence: GNoME has proposed 2.2 million novel crystal structures[7], and OMat24 provided 110 million density-functional-theory calculations spanning diverse compositions[8]. As the datasets have grown, scaling laws—originally identified in natural language processing[9, 10]—have emerged in materials science, showing that predictive accuracy improves with dataset size in a power-law manner, as observed for interatomic potentials,[11] crystal stability,[7] and thermal conductivity[12].

Scaling laws carry a significance beyond empirical curiosity: they turn data acquisition into a quantitative design variable, specifying how much accuracy is gained per increment of data and, conversely, how large a dataset must be to reach a target performance. To date, however, evidence for scaling laws in materials science has come almost exclusively from computationally generated or carefully curated datasets. Materials developed and manufactured in practice differ fundamentally from the idealized systems these datasets describe. Many materials of practical importance—coatings, composites, dispersions, and other multi-component systems—are heterogeneous and often metastable, and their properties are determined not by composition alone but also by processing history, including mixing, dispersing, application, and drying[13–16]. Paints and coatings are an archetypal example: nominally identical formulations can yield films with markedly different microstructures and optical performance depending on the mixing sequence, shear history, and drying conditions. Datasets generated from such systems therefore inevitably contain sample-to-sample variability, particle aggregation, incomplete film formation, and even failed samples. Whether machine learning models trained on such non-ideal experimental data retain the systematic, predictable improvement with dataset size that scaling laws describe remains an open question.

Answering this question is critical for extending MI from computational screening to real-world materials development, particularly in industry. Constructing large experimental datasets requires substantial up-front investment in automation hardware, metrology, and data infrastructure, and—unlike in software or biotechnology—apparatus built for one materials process is rarely transferable to another. A rational decision to invest in data generation therefore requires a quantitative estimate of the return: how much predictive accuracy, and ultimately how much design capability, is gained per additional unit of data. This is precisely the information a scaling law encodes. Establishing whether scaling laws hold for realistic experimental data would thus provide the missing quantitative basis for planning—and justifying—data-generation efforts in materials development.

Testing scaling laws on experimental data requires datasets that are simultaneously large and consistently generated, which has recently become feasible through robotic automation.[17, 18] Automation technologies now cover a wide range of preparation and evaluation tasks, including liquid-handling[19]·[20], solid-state synthesis[21, 22], film fabrication[14, 23], and automated materials characterization.[21–25] Here, we develop an automated experimental system that integrates a pipette robot for liquid-handling[26–28] with infrared (IR) and visible absorption-mapping measurements, enabling the preparation and optical evaluation of paints and coatings at a throughput of up to 1,000 samples per day.

We apply the system to hybrid coatings composed of inorganic pigments dispersed in a polymer matrix for radiative cooling—a passive cooling technology that dissipates heat from the ground through the atmospheric transparency window.[29, 30] While precisely engineered photonic structures can achieve high radiative-cooling performance,[30, 31] paint- and film-based approaches using polymers and pigments have been developed for scalable, large-area radiative cooling.[32–36] Among these approaches, paints are particularly versatile because they can be directly applied to a wide range of substrates. Unlike some film-based radiative-cooling systems that incorporate metallic backing layers,[32, 34] a paint-based approach is particularly attractive when the paint layer itself provides high solar reflectance while maintaining high infrared emissivity.[33, 35, 36] Accordingly, we use

pigment–polymer paints to explore such radiative-cooling performance. Although increasing coating thickness generally improves optical performance,[34] a practical materials-design challenge is to achieve both properties through material composition and coating conditions rather than increased thickness alone. Pigment–polymer paints are also a suitable model system for studying process-dependent materials because their optical properties depend on multiple factors, including composition, particle size and distribution,[32, 35] aggregation,[37] and pores[33] formed during drying, which are in turn governed by paint rheology[38, 39] and droplet geometry on the substrate.[40]

Using this platform, we fabricated and characterized 11,520 coatings across a 9-dimensional space consisting of seven material and two process parameters. Reflecting the process sensitivity of real coatings, the resulting dataset spans wide variations in substrate coverage, hiding power, and particle aggregation—including poorly formed coatings—and we retain this variability in training rather than curating it away. Discriminative neural networks for spectrum-to-parameter and parameter-to-spectrum prediction, together with a spectrum-conditioned diffusion-based generative model (DM), all exhibited power-law scaling of prediction error across training dataset size ranging from 10 to $10^4$ samples, demonstrating that scaling laws established for computational data persist in noisy, process-dependent experimental datasets. The scaling exponents were nearly identical across model architectures and prediction directions, indicating that model performance is governed primarily by the quantity of experimental data. These models also enabled inverse design, in which experimental parameters predicted from a broadband-emitter target spectrum were experimentally validated, with fidelity to the target improving systematically as the training dataset size increased. These results establish scaling laws as a quantitative framework for forecasting the return on experimental data generation, providing a foundation for deploying machine learning in the development of practical, process-dependent materials.

## 2. Results and Discussion

### 2.1 Automated experimental system for radiative-cooling paints and coatings

An automated experimental system for the fabrication and optical measurement of paints and coatings was developed. Fig. 1 shows an overview of the automated experimental system. As shown in the photographs (Figs. 1(a) and 1(b)) and the schematic diagram (Fig. 1(c)), the system consists of three main modules: a robotic unit, a motorized XYZ stage, and an optical evaluation unit (see "Automated experimental equipment" in the Methods section). The robotic unit handles both bottles containing the liquid components and the substrates to be coated and performs liquid aspiration and dispensing via pipetting. The motorized XYZ stage is equipped with a heating table, on which the substrate is placed and precisely positioned. The bottle handling and pipetting mechanics enable the preparation of paints with multiple liquid components at defined ratios. Furthermore, the pipetting operation combined with the XYZ stage enables the formation of 384 coating spots at regularly spaced positions on the substrate. The optical evaluation unit incorporates a Fourier transform infrared spectrometer (FTIR) and a fiber-optic visible-light spectrometer (VIS spectrometer). After the coatings are dried on the heating table, each of the 384 coated spots is sequentially positioned under the optical path by the XYZ stage, enabling the acquisition of IR and VIS spectra for all coatings *via* automated mapping measurement.

In this study, automated experiments were conducted to develop radiative cooling paints and coatings, following the experimental workflow shown in Fig. 1(d). The process comprises three main steps: (i) paint preparation, (ii) drop-casting, and (iii) optical measurement. Detailed descriptions of each process are provided in "Automated experimental workflow" in the Methods section. The individual operations for each process are illustrated in Fig. 1(e). A total of 384 coating spots are fabricated on each substrate. For this purpose, 96 paint mixtures with different compositions, defined by seven material parameters, are prepared in the 96 wells of a well plate. Each mixture is then applied using four different sets of process parameters (DV and DS), yielding four coating spots from each mixture and 384 coating spots in total.

In step (i), the paint mixtures are prepared from seven liquid components prepared in advance: a polyvinylidene fluoride (PVDF)/N,N-dimethylacetamide (DMAc) solution as the matrix, pure DMAc as a diluent, and five pigments with different particle sizes and refractive indices dispersed in DMAc—1 vol% dispersions of $SiO_2$, $CaCO_3$, $BaSO_4$, $Si_3N_4$, and $TiO_2$ (see "Materials Preparation" in the Methods section). The 96 paint mixtures are prepared in batches of 12. For each batch, each liquid component is first homogenized by magnetic stirring and then dispensed into the 12 wells at the prescribed volumes. If the prescribed volume of a liquid component is zero for a given mixture, dispensing of that component into the corresponding well is skipped. This procedure yields 12 paint mixtures with different compositions.

In step (ii), prior to drop-casting, the mixture within each well is mixed by repeated aspiration and ejection using a pipette. Each of the 12 paint mixtures is then drop-cast onto four spots on the substrate using four different sets of process parameters, producing 48 coating spots. This paint-preparation and drop-casting cycle is repeated eight times to produce 96 paint mixtures and 384 coating spots.

The substrate is then dried at 60 °C for 1 h, followed by optical mapping measurements. The fabrication of 384 coating spots and the subsequent optical measurements can be completed in approximately 580 minutes, corresponding to a data production rate of nearly 1,000 measurements per day.

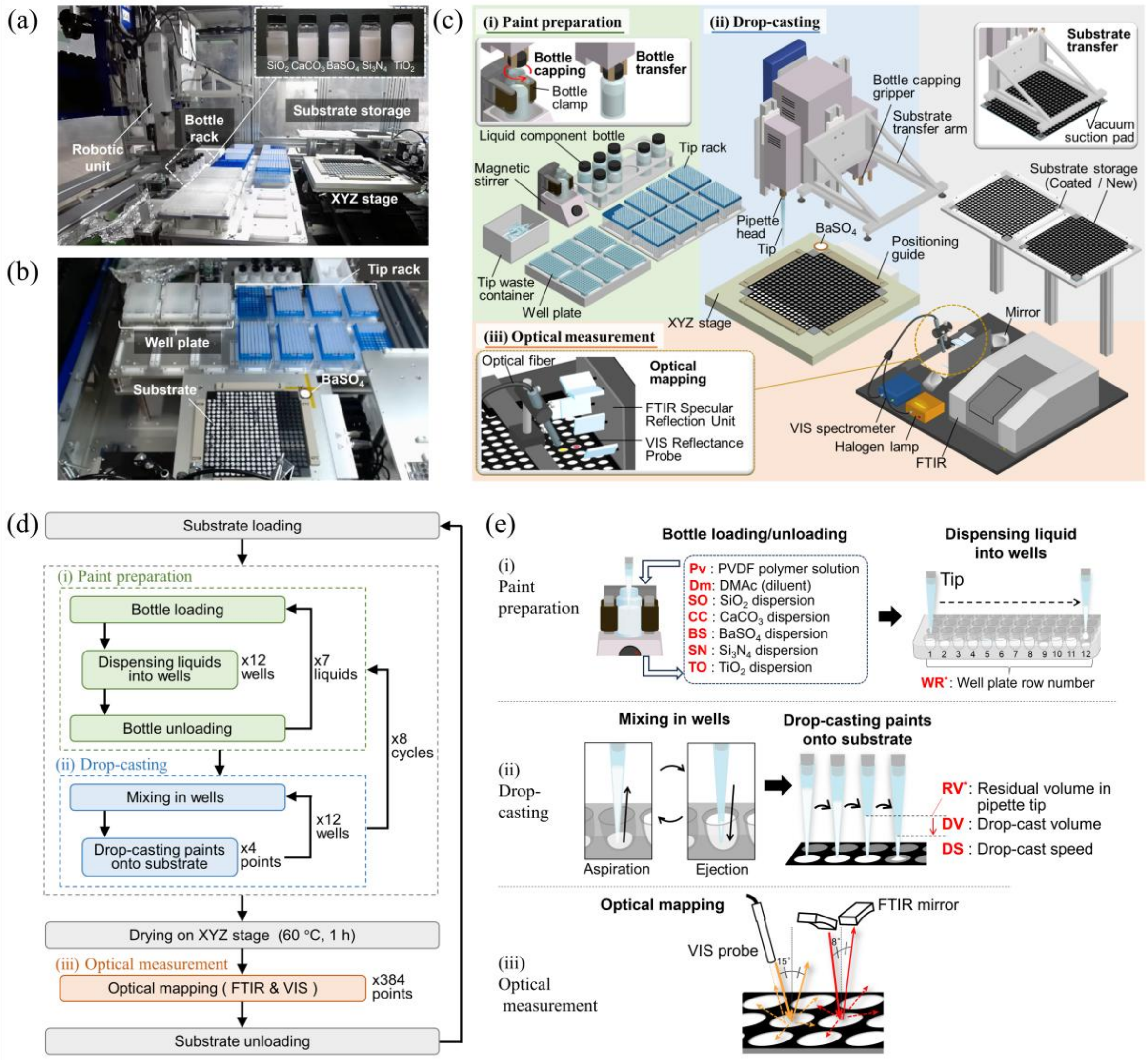

**Fig. 1** Overview of the automated experimental system for high-throughput preparation and optical measurement of radiative-cooling paints and coatings. (a) Side-view photograph of the instrument. Pigment-dispersion bottles are indicated by white dashed lines. (b) Top-view photograph of the instrument. (c) Schematic illustration of the system. Major components associated with the three main processes, (i) paint preparation (green area), (ii) drop-casting (blue area), and (iii) optical measurement (orange area), are highlighted. (d) Experimental workflow illustrating the three-step process. The workflow consists of substrate loading, paint preparation, drop-casting, drying, optical measurement, and substrate unloading. The paint-preparation and drop-casting cycle is repeated eight times to produce 384 coating spots, followed by optical mapping measurements. (e) Detailed schematics of the elemental operations corresponding to the workflow shown in (d): (i) bottle loading/unloading and dispensing liquids into wells; (ii) mixing in wells and drop-casting paints onto substrates; and (iii) optical mapping using FTIR and VIS spectroscopy. The controlled parameters (Pv, Dm, SO, CC, BS, SN, TO, DV, and DS) associated with each operation are shown in red. WR and RV, marked with an asterisk, are passive parameters representing variations that inevitably arise from the sequence of paint-preparation and drop-casting operations. Because these parameters were found to have limited influence on the measured spectra, they were excluded from the set of parameters targeted for optimization in this study.

## 2.2 Experimental operations in a 9-dimensional parameter space

In this workflow, the automated experimental system was used to control a total of nine experimental parameters, consisting of seven material parameters (Dm, Pv, SO, CC, BS, SN, TO) and two process parameters (DV and DS). The parameters associated with each operation are indicated in Fig. 1(e). As material parameters, the system controlled the volumes of seven liquid components used to prepare each paint mixture. The seven material parameters correspond to the volumes of DMAc, PVDF solution, and 1 vol% dispersions of $SiO_2$, $CaCO_3$, $BaSO_4$, $Si_3N_4$, and $TiO_2$ (Fig. 1(e-i); see "Automated Experimental Workflow: (i) Paint preparation" in the Methods section). These material parameters largely determine the infrared absorptivity and visible reflectance of the coatings. The two process parameters were the drop-casting volume (DV) and drop-casting speed (DS) during drop-casting (Fig. 1(e-ii); see "Automated Experimental Workflow: (ii) Drop-casting" in the Methods section). These process parameters influence paint spreading and the resulting coating thickness after drying. A total of 11,520 experimental conditions were randomly generated within the 9-dimensional parameter space. Fig. 2(a) shows a photograph of all 11,520 coatings automatically fabricated on 30 substrates in this study.

In addition to the intended variations introduced through the nine controlled experimental parameters, real experimental datasets inevitably contain non-idealities arising from imperfect uniformity and reproducibility. This can be seen in the inset of Fig. 2(b), where the representative non-uniform samples displayed. Imperfect uniformity can arise from local variations in the dispersion state within the liquid component bottles, well plates, and pipette tips, even after mixing, as well as from local differences in substrate surface cleanliness and wettability or point-to-point variations in rubber-sheet opening positions. Imperfect reproducibility can arise from time-dependent changes in DMAc composition caused by solvent evaporation and moisture uptake due to its hygroscopicity, as well as long-term drift in the optical measurement system. These factors can lead to deviations from nominal compositions, incomplete substrate coverage, particle aggregation, film nonuniformity, and measurement-to-measurement variability. Such unavoidable variations constitute the non-ideal nature of the experimental dataset investigated in this study.

## 2.3 Optical measurement and processing of spectral data

For each substrate, optical mapping measurements were performed on the 384 coatings after drying at 60 °C for 1 h (Fig. 1(e-iii)). FTIR spectra were acquired in transflection mode using the Al-coated substrate as a backing mirror,[41] while visible (VIS) spectra were measured in diffuse-reflection mode (see "Automated Experimental Workflow: (iii) Optical measurement" in the Methods section).

All pigments used in this study had nominal particle sizes below 1 μm (see "Materials Preparation" in the Methods section). Because scattering efficiency is greatest at wavelengths comparable to the particle size,[35] sub-micrometer pigments enhance visible reflectance, which is beneficial for radiative cooling, while they exhibit relatively weak scattering in the infrared region, thereby facilitating accurate emissivity characterization. Nevertheless, residual scattering contributions remain observable in the FTIR-transflection spectra, particularly at wavelengths below 6.5 μm.

FTIR-transflection spectra of the individual pigment samples exhibit an increasing baseline below 6.5 μm, which is attributed primarily to particle scattering. At longer wavelengths, the baseline remains relatively flat but nonzero, reflecting contributions from both transflection-related artifacts and background signals present in the experimental setup.[42, 43] Because diffuse scattering from the coating is not collected by the FTIR optical system, the measured IR spectra reflect attenuation arising from both absorption and scattering processes. To reduce the influence of scattering and clarify the absorption features, baseline correction was applied (see "Spectrum data processing" in the Methods section). Baseline correction was performed by applying asymmetric least-squares (ALS) fitting[44] below 6.5 μm and subtracting a constant background above 6.5 μm. The corrected spectra obtained from all 11,520 measurements are shown in Fig. 2(b). We adopted this relatively simple approach because applying a more rigorous and uniform baseline correction to such a large and diverse experimental dataset is inherently challenging.[45]

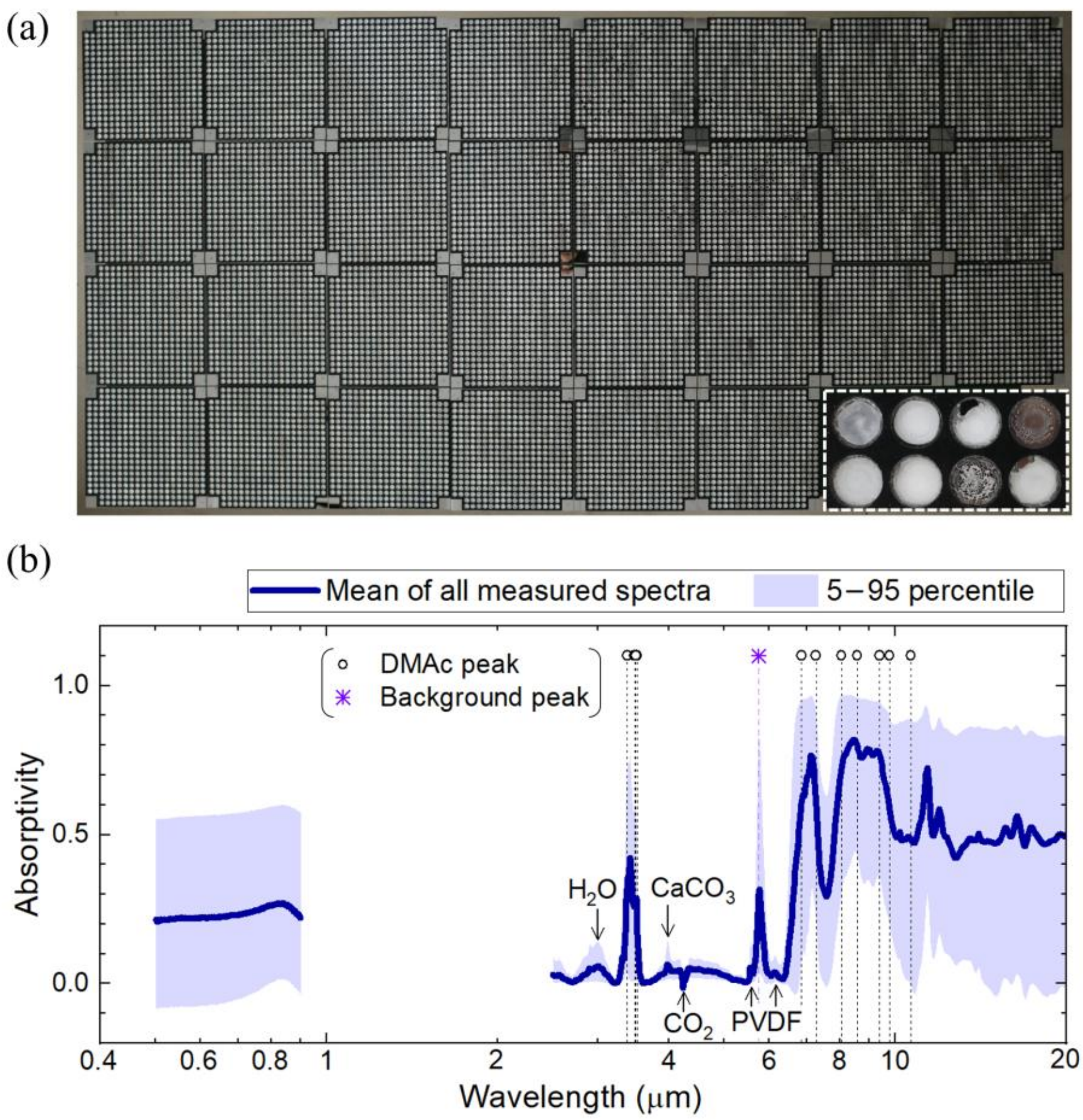

**Fig. 2** Photographs and spectral distribution of the 11,520 coatings fabricated by the automated experimental system. (a) Photograph of 30 substrates, each comprising 384 coatings. The region outlined by the white dashed square highlights representative coating samples. **(b)** The blue line represents the mean spectrum of all fabricated coatings, excluding 27 failed process cases with unphysical properties, i.e. a mean IR reflectivity exceeding 90% over the 2.5–20 μm wavelength range. The light-blue shaded region indicates the 5th–95th percentile range, encompassing 90% of the measured spectra. Major absorption features between 2.5 and 6.5 μm, where residual DMAc and background signals are observed, are identified. Black open circles indicate DMAc peaks, whereas the purple asterisk at 5.75 μm (≈1740 $cm^{-1}$) indicates a background peak. Although the precise origin of the background peak cannot be conclusively assigned, it is likely associated with acrylic-based species leached from the adhesive tape used in substrate assembly. This peak appears consistently in all measured spectra and is therefore regarded as a background peak throughout this study.

## 2.4 Dataset-size scaling of spectrum-to-parameter prediction

Fig. 3 (a) illustrates the machine learning models developed in this study. For spectrum-to-parameter prediction, two complementary approaches were developed and compared: an inverse discriminative neural network (Inverse NN), which deterministically predicts a single parameter set from an input spectrum, and a spectrum-conditioned diffusion-based generative model (DM), which generates multiple candidate parameter sets from the same input spectrum. In addition, a forward discriminative neural network (Forward NN) was developed to predict optical spectra from experimental parameters and to rank the generated parameter sets. Detailed model architectures are provided in the "Machine learning methods" section of the Methods.

Figs. 4(a) and 4(b) show parity plots comparing the experimentally applied parameters with the values predicted by the Inverse NN and the DM, respectively. The training dataset size was systematically varied from 10 to 10,000 samples to evaluate the influence of dataset size on prediction accuracy. For most parameters, the predicted values became increasingly aligned with the corresponding experimental values as the training dataset size increased, demonstrating systematic improvements in model performance. An exception was the drop-casting speed parameter (DS), for which no clear improvement was observed even when 10,000 training samples were used. To systematically investigate the influence of each controlled parameter on the spectral response, SHAP analysis was performed. The results confirm that DS contributes only weakly to the spectral response.

Fig. 4(c) presents the average MAE over all parameters as a function of dataset size, while Fig. 4(d) shows the same relationship on logarithmic axes. For all three models (Inverse NN, Forward NN, and DM), prediction errors decreased systematically as the training dataset size increased. In each case, MAE exhibited an approximately linear relationship with dataset size, indicating power-law scaling between model performance and training data volume.

Power-law relationships between model performance and dataset size are a defining characteristic of scaling laws. Such behavior has been widely reported in natural language processing[10] and has recently been observed in computational materials science, including studies of interatomic potentials,[7, 11, 12] crystal stability, [7] and thermal conductivity prediction.[12] However, these previous studies have primarily relied on computationally generated datasets. In contrast, the present results demonstrate that power-law scaling also emerges in experimentally generated materials datasets obtained through autonomous robotic experimentation. Unlike computational datasets, experimental datasets contain unavoidable variations and noise arising from sample preparation, processing, and measurement. Nevertheless, the observed power-law scaling demonstrates that increasing dataset size remains an effective strategy for improving predictive performance.

The nearly identical scaling exponents obtained for the Inverse NN, Forward NN, and DM suggest that the observed scaling behavior is largely independent of both prediction direction and machine learning architecture. Instead, predictive performance appears to be governed primarily by the quantity and quality of the experimental dataset. These results highlight the importance of large-scale experimental data generation for advancing data-driven materials discovery and inverse materials design.

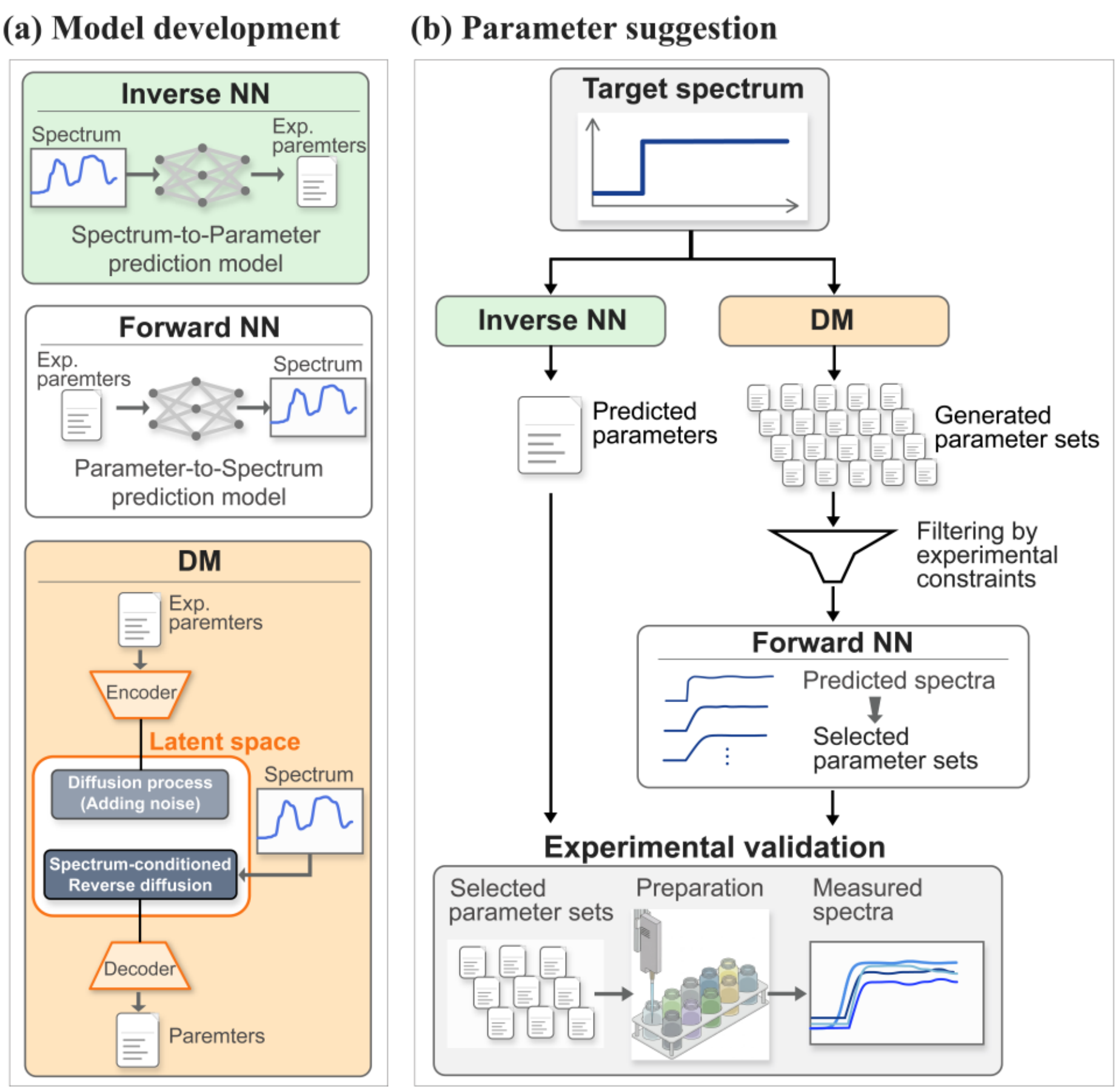

**Fig. 3** Machine learning framework for inverse design. See "Machine-learning methods" in the Methods section for implementation details. (a) Model development. Three machine-learning models were developed using paired optical spectra and experimental parameters. The inverse discriminative neural network (Inverse NN) predicts fabrication parameters from optical spectra, whereas the forward discriminative neural network (Forward NN) predicts optical spectra from fabrication parameters. The spectrum-conditioned diffusion-based generative model (DM) encodes experimental parameters into a latent space, performs forward diffusion (adding noise) and spectrum-conditioned reverse diffusion, and decodes the denoised latent representation into fabrication parameters. (b) Parameter suggestion workflow. Given a target optical spectrum, the Inverse NN directly predicts fabrication parameters, whereas the DM generates multiple candidate parameter sets. After filtering by experimental constraints, the candidates are evaluated using the Forward NN, and the selected parameter sets are experimentally validated.

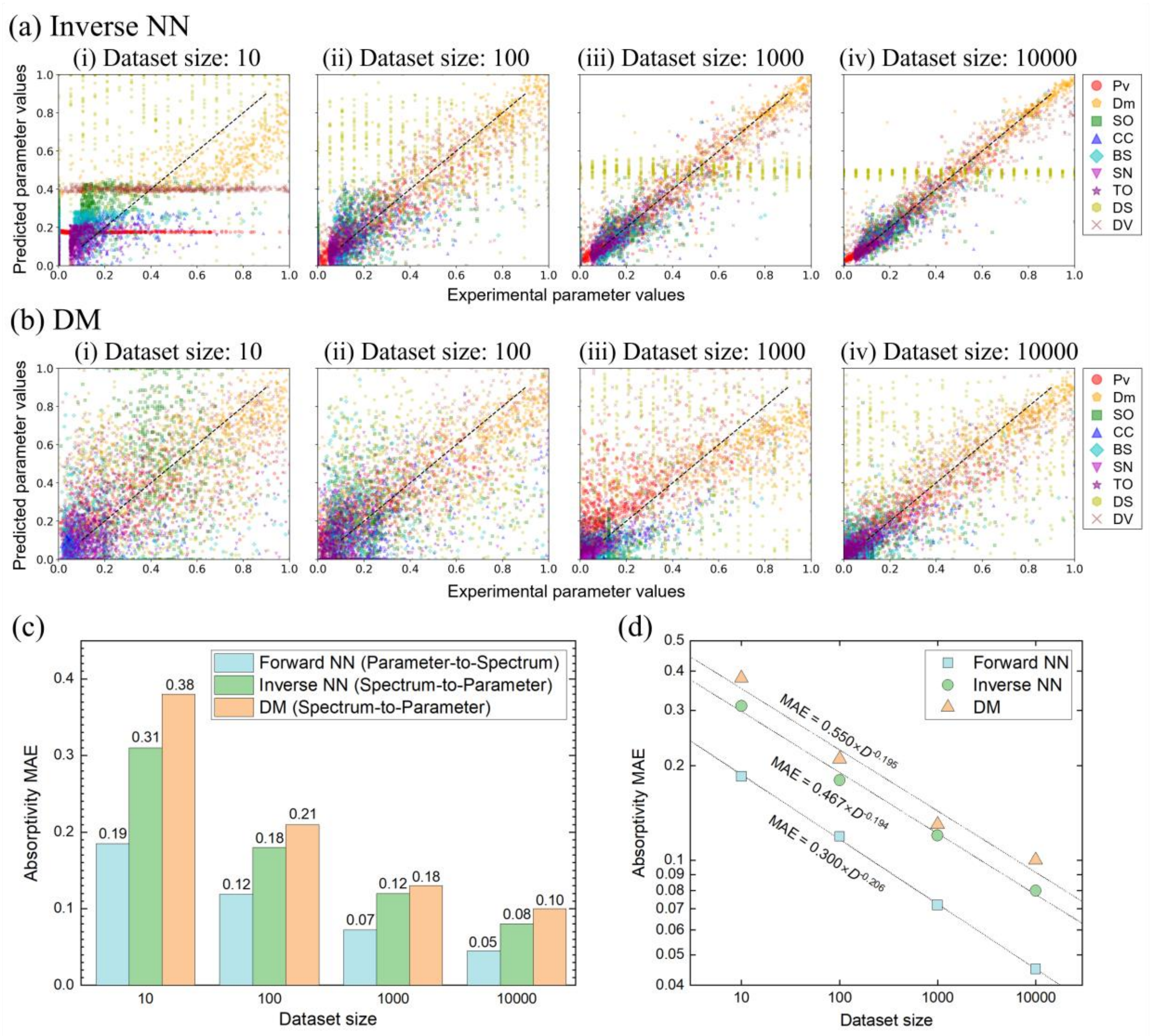

**Fig. 4** Prediction performance of the Inverse NN, Forward NN, and DM. (a) Parity plots of spectrum-to-parameter prediction using the Inverse NN trained with dataset sizes of 10, 100, 1,000, and 10,000. Parity plots for the nine coating preparation parameters (Pv, Dm, SO, CC, BS, SN, TO, DS, and DV) are overlaid in each panel. The horizontal and vertical axes represent the experimental and predicted parameter values, respectively. To improve visualization, both experimental and predicted parameter values were normalized using min–max scaling. (b) Parity plots of spectrum-to-parameter prediction using the DM trained with varying dataset sizes. (c) Mean absolute errors (MAEs) of the Forward NN, Inverse NN, and DM as functions of training dataset size. (d) Relationship between training dataset size and MAE shown on log-log axes. The prediction error decreases approximately linearly on the log–log plot, indicating power-law scaling between model performance and dataset size. Here, D is the training dataset size used in the power-law fitting.

## 2.5 Experimental validation of inverse design based on predicted parameters

To evaluate the practical inverse-design capability of the developed machine learning models, experimental parameters were predicted from a target optical spectrum and experimentally validated. Fig. 3(b) illustrates the parameter-suggestion workflow used for inverse design. The Inverse NN directly predicts a single experimental parameter set, whereas the diffusion-based generative model (DM) generates multiple parameter sets for the same target spectrum. These generated parameter sets are filtered according to experimental constraints, ranked using the Forward NN, and experimentally validated.

In this study, a broadband-emitter target spectrum was selected to demonstrate the inverse-design workflow. Radiative-cooling performance is determined by the emissivity spectrum of a coating. Broadband emitters exhibit high emissivity over a broad infrared wavelength range, thereby maximizing thermal radiation.[46] According to Kirchhoff's law, emissivity is equal to absorptivity at thermal equilibrium. Therefore, a target absorptivity spectrum was defined and used as the design objective. The threshold wavelength of 6 μm was selected because the baseline-corrected spectra exhibit most of their variation above this wavelength, whereas spectral variation below 6 μm is largely suppressed by the baseline-correction procedure. Accordingly, the target spectrum was defined over the wavelength range containing the majority of the spectral information in the present dataset.

The trained Inverse NN and DM were then applied to the target spectrum, and the resulting parameter sets were experimentally evaluated by fabricating coatings and measuring the resulting optical spectra. The similarity between the target and measured spectra was quantified using the spectral difference defined by equation (1).

$$\begin{aligned}\text{Spectral difference} = & \frac{1}{\lambda_{\mathrm{VIS,\,max}} - \lambda_{\mathrm{VIS,\,min}}} \sum_{i=1}^{N_{\mathrm{VIS}}} \left| A_{\mathrm{sample}}\left(\lambda_{\mathrm{VIS},\,i}\right) \text{-} A_{\mathrm{target}}\left(\lambda_{\mathrm{VIS},\,i}\right) \right| \Delta\lambda_{\mathrm{VIS},\,i} \\ & + \frac{1}{\lambda_{\mathrm{IR,\,max}} - \lambda_{\mathrm{IR,\,min}}} \sum_{i=1}^{N_{\mathrm{IR}}} \left| A_{\mathrm{sample}}\left(\lambda_{\mathrm{IR},\,i}\right) \text{-} A_{\mathrm{target}}\left(\lambda_{\mathrm{IR},\,i}\right) \right| \Delta\lambda_{\mathrm{IR},\,i}\end{aligned} \tag{1}$$

where $N_{\mathrm{VIS}}$ and $N_{\mathrm{IR}}$ are the numbers of wavelength data points in the VIS (0.5–0.9 μm) and IR (2.5–20 μm) wavelength ranges, respectively; $\lambda_{\mathrm{VIS},\,i}$ and $\lambda_{\mathrm{IR},\,i}$ denote the corresponding wavelengths; and $\Delta\lambda_{\mathrm{VIS},\,i}$ and $\Delta\lambda_{\mathrm{IR},\,i}$ are the wavelength intervals associated with each data point.

Fig. 5(a) compares the target spectrum with the sample with the lowest total spectral difference among the original 11,520 experimentally measured coatings. Figs. 5(b) and 5(d) show the spectra obtained using the parameter sets suggested by the Inverse NN and the DM, respectively. As the training dataset size increased, the experimentally measured spectra progressively approached the target spectrum. The total spectral difference and the contributions from the VIS and IR wavelength ranges are shown separately in Figs. 5(c) and 5(e). Although the improvement is less smooth than the scaling behavior shown in Fig. 4 because fabrication and measurement variability are included, the same overall trend is observed. Models trained on larger datasets generally generated parameter sets that produced spectra closer to the target.

These results demonstrate that the scaling behavior observed in Fig. 4 directly translates into improved inverse-design performance. Larger experimental datasets produce more accurate experimental parameter predictions, resulting in coatings whose optical spectra more closely reproduce the desired target spectrum. The developed framework therefore enables experimental parameters to be designed directly from target optical spectra, providing a practical route for tailoring spectral properties through data-driven inverse design.

Although the inverse-designed coatings did not surpass the best sample already present in the original dataset of 11,520 samples, the developed framework successfully identified experimental parameter sets that reproduced the target spectrum with high fidelity within the current experimental parameter space. The largest remaining mismatch was in the VIS range, where improvement with increasing dataset size was limited. This mismatch may also reflect the limited angular range of diffuse light captured by the VIS measurement setup, as well as the sensitivity of visible-light performance to coating surface morphology and uniformity.

Many reported high-performance radiative-cooling paints without a back reflector have coating thicknesses of around 100–500 μm or more.[33, 35, 36] By contrast, assuming uniform coverage and a dense film, the solid contents and dispensed volumes used in this study correspond to an estimated thickness of 8–37 μm for the inverse-designed coatings. Although their actual thickness was not measured and may vary with paint spreading and porosity, even assuming a porosity of approximately 50%, a value reported for a PVDF-based cast film with deliberately introduced pores,[33] the estimated thickness remains below 100 μm. This thickness range is comparable to that of typical industrial and architectural finish coats. The present results demonstrate the feasibility of optimizing paint composition and coating conditions in this practically relevant thickness range.

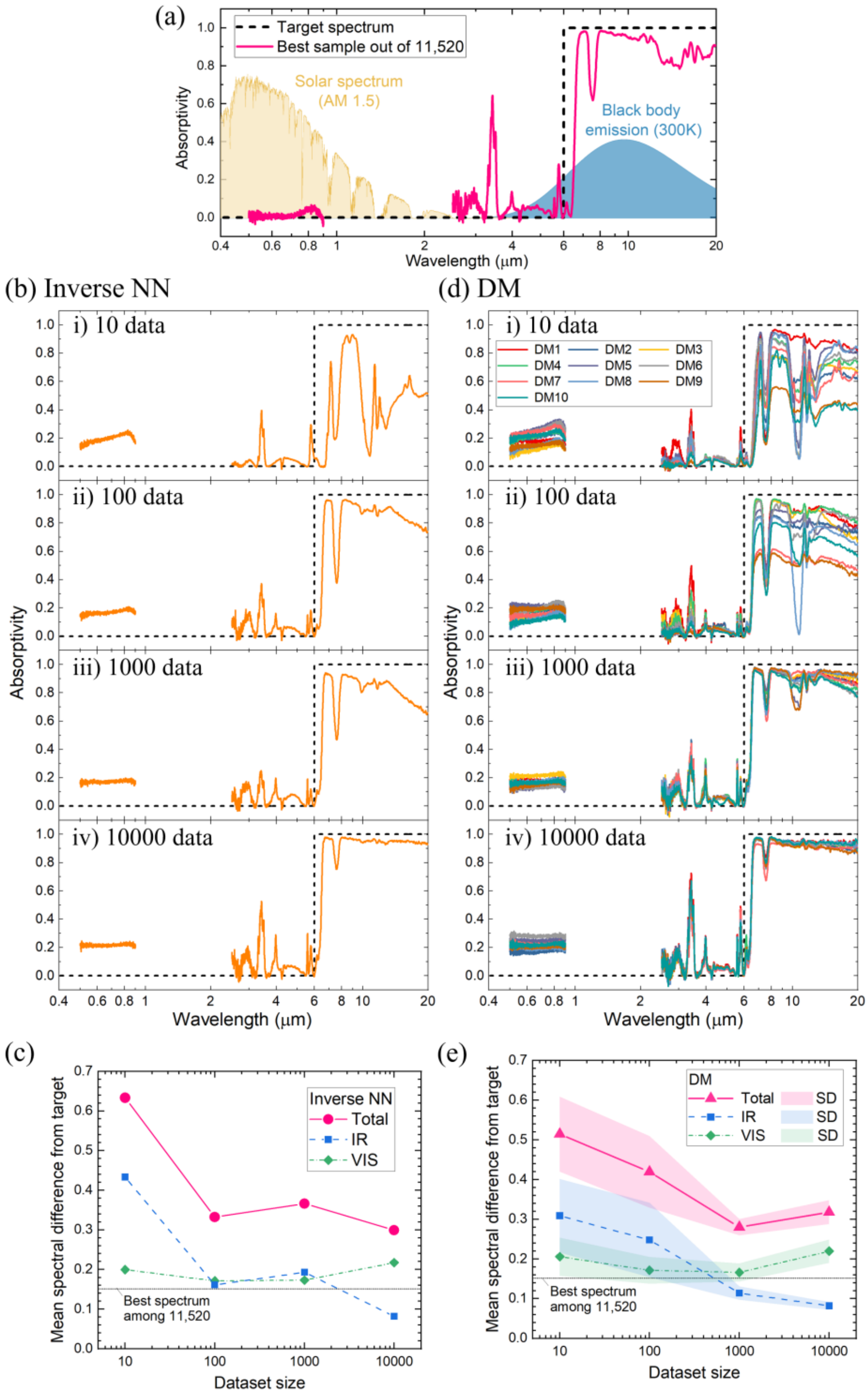

**Fig. 5** Experimental validation of inverse design for a broadband-emitter target spectrum. (a) Target absorptivity spectrum (black dashed line), defined as an absorptivity of 1 above 6 μm and 0 below 6 μm. The pink line indicates the sample with the lowest total spectral difference among the 11,520 measured coatings. For reference, the AM1.5G solar spectrum and the blackbody emission spectrum at 300 K are shown in orange and blue, respectively. The solar spectrum was taken from the ASTM G173-03 reference spectrum.[47] The blackbody radiation spectrum at 300 K was calculated using Planck's law. (b) Experimental spectra obtained using parameter sets suggested by the inverse NN trained with dataset sizes of (i) 10, (ii) 100, (iii) 1,000, and (iv) 10,000 samples. (c) Mean total spectral difference

and the contributions from the VIS and IR wavelength ranges between the target spectrum and the experimentally measured spectra obtained using the Inverse NN. The dotted line indicates the sample with the lowest total spectral difference among the original 11,520 coatings. (d) Experimental spectra obtained using ten parameter sets proposed by the DM (DM1–DM10). Models were trained with dataset sizes of (i) 10, (ii) 100, (iii) 1,000, and (iv) 10,000 samples. (e) Mean total spectral difference and the contributions from the VIS and IR wavelength ranges between the target spectrum and the experimentally measured spectra obtained using the DM. The solid lines indicate the means, and the shaded regions represent the standard deviations (SDs). The dotted line indicates the spectral difference of the sample with the lowest total spectral difference among the original 11,520 coatings. Spectral differences were calculated using equation (1).

## 2.6 Interpretation of parameter importance and the effective parameter space

Although the DM exhibited slightly larger prediction errors than the Inverse NN, both models showed nearly identical scaling exponents (Fig. 4). This result suggests that the benefit of increasing dataset size is largely independent of the machine learning architecture. Because the mapping from experimental parameters to optical spectra is many-to-one, the inverse problem is intrinsically ill-posed. The Inverse NN provides a single deterministic parameter estimate, whereas the DM samples from a conditional distribution of feasible parameter combinations. In the low-data regime, the Inverse NN predictions collapsed toward restricted parameter ranges, whereas the DM retained broader coverage of the feasible parameter space. This suggests that the generative formulation can mitigate mode collapse in inverse prediction and provide multiple feasible parameter sets, although its stochastic nature also leads to larger prediction variance. Such diversity is advantageous for experimental design because it provides alternative parameter sets that can be selected according to practical constraints such as material availability, process robustness, or cost, rather than being restricted to a single deterministic prediction. The generated parameter distributions therefore provide insight into the effective parameter space learned from the experimental dataset.

Fig. 6(a) shows the parameter distributions generated by the DM for the target spectrum. Several parameters are concentrated within relatively narrow ranges, whereas others remain broadly distributed, indicating that some parameters are more strongly constrained than others for achieving the target spectrum.[48] Fig. 6(b) further compares the experimental parameter set predicted by the Inverse NN with the multiple parameter sets generated by the DM. Although the DM generates multiple parameter sets, many occupy similar regions of the experimental parameter space, suggesting that only a limited number of effective design directions exist within the current parameter space.

For the broadband-emitter target, the generated parameter sets consistently converged toward more concentrated paint compositions with relatively low DMAc contents and high PVDF contents. This tendency may partly reflect the spectrum preprocessing adopted in the present study. Because the baseline-correction procedure is performed in the absorbance domain (see "Spectrum data processing" in the Methods section), part of the observed convergence toward low-DMAc compositions may reflect preprocessing artifacts in addition to intrinsic material characteristics. Although these tendencies should not be interpreted as causal relationships without further mechanistic validation, they illustrate that conditional generative models can provide interpretable design hypotheses in addition to feasible parameter sets.

The predictive and generative models developed here are expected to be most reliable for interpolation within the experimental domain covered by the training data. Their extrapolation capability to unseen material chemistries, substantially different composition ranges, or untested processing conditions remains limited. In particular, the DM tends to sample parameter sets from the learned feasible distribution and therefore should not be regarded as an unrestricted generator outside the experimentally explored design space. Furthermore, the present study is limited to a single polymer matrix, five inorganic pigments, and a drop-casting-based fabrication process. Consequently, the learned design rules may not be directly transferable to other coating chemistries or processing routes. Future work should expand both the material and process spaces, incorporate uncertainty quantification and out-of-distribution detection, and implement closed-loop active learning to improve both interpolation and extrapolation performance.

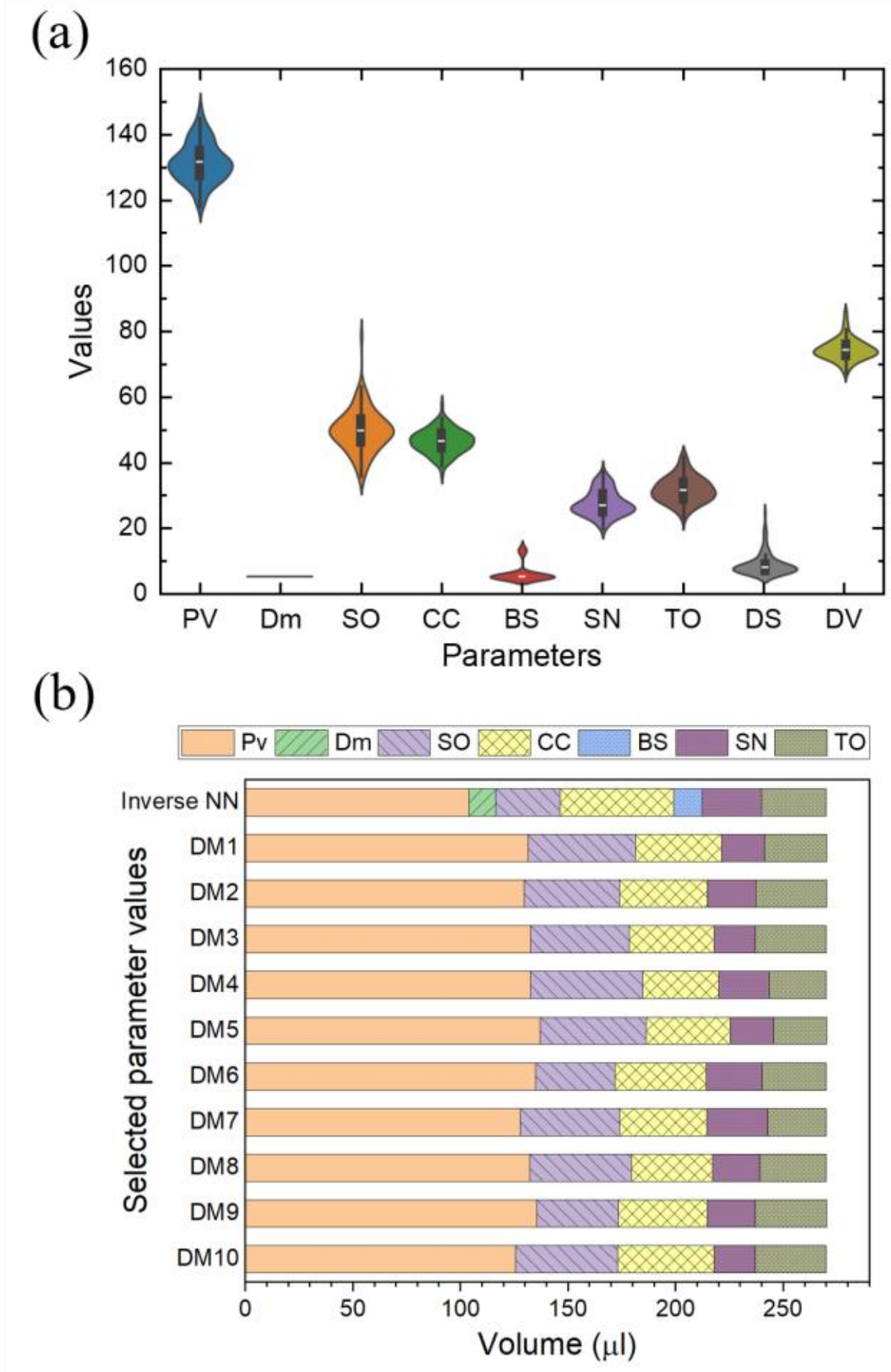

**Fig. 6** Interpretation of parameter importance and the effective parameter space based on DM-generated parameter distributions. (a) Violin plots showing the distributions of parameter values suggested by the DM for achieving the target spectrum. The DM was trained using a dataset of 10,000 samples. The width of each violin represents the probability density of the suggested values. Narrow distributions indicate parameters that are more strongly constrained for achieving the target spectrum, whereas broad distributions indicate greater flexibility. (b) Parameter values predicted by the Inverse NN and generated by the DM for the target spectrum. Results are shown for a training dataset size of 10,000. Ten parameter sets (DM1–DM10) were selected from the candidate parameter sets generated by the DM.

## 3. Conclusion

In this study, we developed an automated experimental system for the preparation and optical evaluation of radiative-cooling paints and coatings. Using this system, 11,520 coating samples were fabricated and characterized within a 9-dimensional parameter space consisting of seven material parameters and two process parameters.

The Inverse NN, Forward NN, and DM were trained using subsets of the experimental dataset containing 10, 100, 1,000, and 10,000 samples. Prediction errors decreased systematically with increasing training dataset size, following a power-law relationship. Importantly, this scaling behavior persisted despite the non-ideal variations inherent to the experimentally generated dataset. Larger datasets also improved inverse-design performance, enabling experimental parameters to be predicted from target optical spectra. These results demonstrate that increasing dataset size remains an effective strategy for improving machine-learning performance even for non-ideal experimental data.

Although the DM exhibited larger prediction errors than the Inverse NN, both models showed nearly identical scaling exponents. This result suggests that scaling behavior is governed primarily by the quality and quantity of the experimental dataset rather than by the machine learning architecture. In addition, the DM generated multiple feasible parameter sets for a given target spectrum, and analysis of the resulting parameter distributions provided insight into parameter importance and the effective parameter space.

These results demonstrate the potential of combining autonomous experimentation and machine learning for the development of radiative-cooling paints and coatings. The automated workflow enabled the generation of large experimental datasets, while the resulting data were used to investigate scaling behavior, guide inverse design, and quantitatively interpret the effective parameter space. The proposed framework provides a practical route toward data-driven inverse design of radiative-cooling coatings and other process-dependent materials.

## 4. Methods

### 4.1 Automated experimental equipment

An automated experimental system consists of a robotic unit, a motorized XYZ stage, and a spectroscopy unit. The robotic unit (customized PIPETTE MASTER, Musashi Engineering) comprises a Cartesian robot equipped with a pipette head, a bottle-capping gripper and a substrate-transfer arm. This robotic unit is responsible for (i) paint preparation and (ii) drop-casting (Figs. 1(d-i) and 1(d-ii)). The motorized XYZ stage accommodates the substrate on the table and moves it during (iii) the optical measurement process. The table was maintained at 60 °C *via* resistive heating throughout the experiment to ensure the drying of the coatings under constant conditions. The spectroscopy unit consists of an FTIR (IRTracer-100, Shimadzu) and a fiber-optic visible-light spectrometer (VIS spectrometer).

### 4.2 Materials preparation

Polyvinylidene fluoride (PVDF) / N,N-dimethylacetamide (DMAc) solution as the polymer matrix, and 1 vol% DMAc dispersions of $SiO_2$, $CaCO_3$, $BaSO_4$, $Si_3N_4$, and $TiO_2$ as inorganic pigments were prepared. 31.0 g of PVDF (ARKEMA, Kynar Flex® 2801-00, $\rho$ = 1.8 g cm$^{-3}$) was dissolved in 400 mL of DMAc (FUJIFILM Wako Pure Chemical Corporation, 98.0+%) by heating to 60 °C for 3 h. Dispersions were prepared by adding 4.5 g of $SiO_2$ (NIPPON SHOKUBAI, SEAHOSTAR® KE-S50, $d$ = 0.5 μm), 5.5 g of $CaCO_3$ (BIHOKU FUNKA KOGYO, Softon3200, $d$ = 0.7 μm), 9.1 g of $BaSO_4$ (Takehara Kagaku Kogyo, TS-3, $d$ = 0.6 μm), 6.4 g of $Si_3N_4$ (E M Japan, NP-SI3N4-3, $d$ = 0.8 μm), and 8.5 g of $TiO_2$ (E M Japan, NP-TIO2-14, $d$ = 0.5 μm, rutile phase) to 200 mL of DMAc, followed by stirring with a magnetic stirrer for several days. The resulting solution and dispersions were aliquoted into 28 mL vials and loaded into the system.

### 4.3 Labware used in the automated experimental system

The liquid components were stored in 28 mL borosilicate glass screw bottles and loaded into the system. A magnetic stir bar was placed in each bottle for stirring dispersions. 300 μL polypropylene pipette tips (505C, WATSON) were used for pipetting, and 96-well polypropylene plates (270 μL, Porvair) were used for preparing paint mixtures. The substrate containing 384 coating spots was prepared by bonding a rubber sheet to a glass substrate. 200 × 200 mm Tempax glass (1 mm thick) was used as the base material. Ethylene propylene terpolymer (EPT) rubber sheet (1 mm thick) was processed using a water cutter to create 384 holes with a diameter of 8.6 mm. It was affixed to the glass substrate using double-sided tape. The surface of the glass substrate was coated with an Al film via sputtering. The Al film served as a backing mirror for the coating during optical measurements, enabling FTIR transflection measurements.

#### 4.4 Automated experimental workflow

**(i) Paint preparation**

The paint preparation followed a three-step procedure: bottle loading, liquid dispensing, and bottle unloading (Figs. 1(d), 1(e-i)). First, the liquid component bottle was transferred from the bottle rack to a magnetic stirrer using a bottle-capping gripper mounted on the robotic unit (bottle loading step). While the bottles were stored in the rack, their screw caps were kept closed to prevent solvent evaporation. Once placed on the stirrer, the bottle was secured using a bottle clamp, and the gripper was rotated to uncap it. If the bottle contained a dispersion, it was stirred for 4 s using the magnetic stirrer to ensure uniformity.

A pipette tip was then inserted into the liquid, and a predetermined volume was aspirated and dispensed into a well. This process was repeated for 12 wells corresponding to one column of the well plate (liquid dispensing step). If the prescribed volume of a liquid component was zero for a given mixture, dispensing of that component into the corresponding well was skipped. After completing the dispensing for one column, the bottle was recapped and returned to the bottle rack from the magnetic stirrer (bottle unloading step). Subsequently, the next liquid component bottle was placed on the magnetic stirrer, and the same procedure was repeated. This three-step process was performed for each of the seven liquid components, producing one batch of 12 paint mixtures with different compositions. The composition of each mixture was defined by seven material parameters: Dm, Pv, SO, CC, BS, SN, and TO.

### (ii) Drop-casting

Drop-casting followed a two-step procedure: liquid mixing and mixture drop-casting (Figs. 1(d), 1(e-ii)). The mixture in each well was mixed by performing 8 cycles of aspiration and ejection using a pipette tip (mixing step). Each mixture was then drop-cast onto four spots on the substrate using four different sets of process parameters, namely the drop-casting volume (DV) and speed (DS) (drop-casting step).

After drop-casting 48 spots from the 12 prepared mixtures, the next batch of 12 mixtures was prepared and subsequently applied to another 48 spots. This paint-preparation and drop-casting cycle was repeated eight times, resulting in 96 paint mixtures and 384 drop-cast spots. The deposited paints were dried at 60 °C for 1 h to form 384 coatings.

### (iii) Optical measurement

FTIR measurements were performed in transflection mode using a specular reflection optical system (SRM-8000, Shimadzu). A collimated IR beam was extracted from the IRTracer-100, focused by an ellipsoidal mirror, and directed into the SRM-8000. The SRM-8000 was installed in an inverted configuration, enabling reflectance measurements on samples positioned beneath the optical system with an average incidence angle of 10 °. The aperture was adjusted to set the illumination spot size on the sample coating to $\Phi$5 mm. The specularly reflected light from the sample was collected by another ellipsoidal mirror, which focused it onto a Deuterated L-Alanine Triglycine Sulphate (DLATGS) detector. An exposed Al mirror area was used as a reflectance standard for IR measurement. The measurement conditions were set to five scans, a wavenumber range of 350–7800 cm$^{-1}$, and a mirror speed of 2.8 mm s$^{-1}$. However, the actual analysis was limited to the 500–4000 cm$^{-1}$ range. This restriction was applied because the high-wavenumber region exhibited significant noise, and the 448 cm$^{-1}$ region showed peaks originating from optical system vibrations and environmental disturbances, necessitating its exclusion from the dataset used for analysis.

VIS measurements were performed in diffuse reflection mode using a bifurcated probe (R600-8-UV-VIS-SR,

StellarNet) for reflectance measurements. The probe has coaxially bundled illumination and detection fibers. Light from a halogen light source (SL1, StellarNet) was delivered to the sample via the illumination fiber, and the reflected light from the sample was introduced into the spectrometer (Blue-Wave VIS, StellarNet) via the detection fiber. The probe was set at a 15° incidence angle to evaluate the reflectance properties of the coating without being affected by specular reflection from the Al mirror on the rear side of the coating. In normal 0° incidence measurements, specular reflection from the Al mirror can be directly detected if the hiding power of the coating is insufficient. Employing a 15° incidence angle ensures that only reflection from the coating is detected by capturing diffuse reflection returning in the incident direction. The illumination spot size on the sample coating was Φ5 mm with a sample-probe distance of 6.5 mm. For VIS spectroscopy, a compressed $BaSO_4$ powder (Sigma-Aldrich, 99.99%) mounted on the XYZ stage was used as the reflectance standard.

## 4.5 Coating preparation parameters and dataset generation

Dm, Pv, SO, CC, BS, SN, and TO represent the volumes of DMAc solvent, PVDF polymer solution, $SiO_2$, $CaCO_3$, $BaSO_4$, $Si_3N_4$, and $TiO_2$ dispersions, respectively. To ensure proper droplet detachment during dispensing, the minimum volume of each liquid component was set to 6.5 μL. Each mixture contained PVDF as the polymer matrix together with at least one pigment, while unused pigments were assigned a volume of 0 μL. The dispensing volume was controlled with a precision of 0.1 μL. In addition, a total volume constraint of 270 μL was imposed: Dm + Pv + SO + CC + BS + SN + TO = 270. Each liquid component was sequentially dispensed into a single column (12 wells) of the well plate. First, DMAc was dispensed into all 12 wells, followed by PVDF, $SiO_2$, $CaCO_3$, $BaSO_4$, $Si_3N_4$, and $TiO_2$ in the same manner (Fig. 1(e-i)). As a result, each well contained a paint mixture with a total volume of 270 μL.

In the drop-casting process, each paint mixture was applied to the substrate through four consecutive drop-casting steps after mixing by eight cycles of aspiration and ejection. During each drop-casting step, the drop-casting volume (DV) and drop-casting speed (DS) were controlled. Because drop-casting was performed sequentially across the 12 wells within a column, the well row number (WR) was also considered as a parameter representing the coating order and the elapsed time before deposition. In addition, the remaining volume in the pipette tip at the beginning of each drop-casting step (RV) was considered as a parameter because the amount of liquid remaining in the tip varied among the four consecutive drop-casting steps. Because excessively thick coatings can attenuate the IR signal in the transflection geometry, the ranges of the composition and drop-casting parameters were set to keep the resulting coatings within the measurable range of the setup.

After coating 384 spots on a substrate, the deposited paints were dried at 60 °C for 1 h, followed by optical measurements. The table temperature was maintained at 60 °C throughout both the drop-casting and optical measurement processes.

WR and RV were initially included as candidate parameters. However, SHAP analysis showed that their contributions to the measured spectra were limited. Therefore, WR and RV were excluded from model training, and Dm, Pv, SO, CC, BS, SN, TO, DV, and DS were used as the nine experimental parameters for machine learning analysis. A total of 11,520 experimental parameter sets were randomly generated by varying these nine parameters.

## 4.6 Spectrum data processing

In FTIR measurements, the detected signal corresponds to IR light that passes twice through the coating after being specularly reflected by the rear-side Al mirror. Since the optical system is designed for specular reflection, diffuse scattering is not detected. However, the measured spectrum represents an apparent absorption spectrum because light attenuation occurs due not only to intrinsic absorption within the coating, but also to light extinction caused by scattering. To reduce the influence of scattering and clarify the absorption-related spectral features, baseline correction was applied. The correction was performed by first converting the spectrum to the absorbance scale, which allows the additive treatment of absorption and scattering components. Then, in the wavelength range of 2.5–6.5 μm, baseline fitting was conducted using the Asymmetric Least Squares (AsLS) method. The baseline was obtained by minimizing the following objective function: $S = \sum_i w_i(y_i - z_i)^2 + \lambda\sum_i(\Delta^2 z_i)^2$. Here, $y_i$ is the original spectrum, $z_i$ is the estimated baseline, $w_i$ is the weight for each point, and $\lambda$ is the smoothing parameter. The resulting baseline curve was subtracted from the measured spectrum to suppress the scattering contribution. Although absorption peaks originating from residual DMAc solvent and background impurities are observed in this region, the baseline contribution was assumed to arise primarily from scattering. For the long-wavelength region above 6.5 μm, the baseline value at 6.5 μm was applied uniformly as a constant approximation. Finally, the corrected spectrum was converted to absorptance and used for emissivity analysis. In contrast, in the visible region, measurements were performed at a 15 ° incidence angle to capture diffuse reflection caused by scattering at the pigment–matrix interface and the coating surface. In this case, no baseline correction was applied, and the reflectance data were used directly for further analysis.

## 4.7 Machine learning methods

Two multilayer-perceptron-based discriminative neural networks were developed: an inverse discriminative neural network (Inverse NN) for spectrum-to-parameter prediction and a forward discriminative neural network (Forward NN) for parameter-to-spectrum prediction. In addition, a spectrum-conditioned diffusion-based generative model (DM) was developed to generate multiple feasible parameter sets corresponding to a target spectrum. For each training dataset size (10, 100, 1,000, and 10,000), the training samples were selected from the experimental dataset. During inverse design, the generated parameter sets were filtered according to experimental constraints, evaluated using the Forward NN, and ranked by their spectral deviation from the target. The top-ranked parameter sets were selected for experimental validation.

#### (i) Inverse discriminative neural network (Inverse NN) for spectrum-to-parameter prediction

We used the Inverse NN to map an optical spectrum to the nine experimental parameters. The implemented architecture consisted of three one-dimensional convolutional layers with 16, 32 and 64 channels, kernel size 5 and max-pooling after each convolution. The flattened representation is passed through fully connected layers of 512, 256 and 128 units with ReLU activations and dropout of 0.3, followed by a nine-dimensional output layer. The model was trained with mean-squared error and an additional constraint penalty weighted by $1\times10^{-4}$.

**(ii) Forward discriminative neural network (Forward NN) for parameter-to-spectrum prediction**

We used the Forward NN to map the nine experimental parameters to an optical spectrum and to rapidly evaluate and rank the parameter sets generated by the DM before experimental validation. The implemented architecture consists of fully connected layers with 128, 256, 512, 1024 and 2048 hidden units, each followed by batch normalization, ReLU activation and dropout of 0.3, and a final tanh output layer. The training loss combined mean-squared error, cosine spectral similarity and total-variation regularization to promote accurate and smooth predicted spectra.

**(iii) Spectrum-conditioned diffusion-based generative model (DM) for spectrum-to-parameter prediction**

We developed a spectrum-conditioned diffusion-based generative model (DM) for spectrum-to-parameter inverse design by extending the tabular-data diffusion strategy used in TABSYN.[49, 50] Experimental parameters were represented as tabular data, and each normalized nine-dimensional experimental parameter vector was treated as a single tabular record and encoded by a variational autoencoder (VAE) into a continuous latent representation. Conducting diffusion in this latent parameter space enables the model to sample from a compact and smooth manifold of experimentally feasible parameter combinations, rather than directly modeling the original parameter space.

To enable spectrum-guided inverse design, the target or measured absorptivity spectrum was also encoded into a compact spectral embedding. This embedding was provided to the denoising network during diffusion sampling, together with the noisy latent parameter vector and the diffusion time step. Consequently, the reverse diffusion process is conditioned on the input spectrum and learns to generate parameter candidates that are both consistent with the learned distribution of experimental parameters and relevant to the desired spectral response. The generated latent samples were subsequently decoded by the VAE decoder and inverse-normalized to obtain experimental parameter values, which were then subjected to constraint filtering and spectral evaluation.

## Acknowledgements

This research was supported by Japan Science and Technology Agency (JST) CREST under the projects "Exploring Innovative Materials in Unknown Search Space" (JPMJCR21O2).

## Author Contribution Statement

K.S., K.T. and J.S. conceived and designed the study. K.S., Y.L., Y.A., and T.L. developed the automated experimental system and performed the experiments. K.S. collected and analyzed the experimental data. J.L. and K.T. developed the machine-learning models and performed the machine-learning analyses. K.S. and J.L. analyzed the results and drafted the manuscript with input from all authors. All authors discussed the results and commented on the manuscript. J.S supervised the project.

## Competing interests

The authors have no competing interests to declare that are relevant to the content of this article.